\documentclass[12pt,a4paper]{article}
\usepackage{epsfig}
\usepackage{axodraw}
\def\nl{\nonumber\\}
\def\beq{\begin{equation}}
\def\eeq{\end{equation}}
\def\beqar{\begin{eqnarray}}
\def\eeqar{\end{eqnarray}}
\def\bfi{\begin{figure}}
\def\efi{\end{figure}}
\def\btab{\begin{table}}
\def\etab{\end{table}}
\def\bce{\begin{center}}
\def\ece{\end{center}}
\def\bit{\begin{itemize}}
\def\eit{\end{itemize}}

\def\scrs{\scriptscriptstyle}
\def\text{\textstyle}

\def\al{\alpha}

\def\de{\delta}

\def\la{\lambda}
\def\si{\sigma}

\def\refeq#1{\mbox{(\ref{#1})}}
\def\reffi#1{\mbox{Fig.~\ref{#1}}}

\def\refse#1{\mbox{Sect.~\ref{#1}}}

\def\citere#1{\mbox{Ref.~\cite{#1}}}
\def\citeres#1{\mbox{Refs.~\cite{#1}}}

\newcommand{\GeV}{\unskip\,\mathrm{GeV}}

\newcommand{\TeV}{\unskip\,\mathrm{TeV}}

\def\mathswitchr#1{\relax\ifmmode{\mathrm{#1}}\else$\mathrm{#1}$\fi}

\newcommand{\PW}{\mathswitchr W}

\def\mathswitch#1{\relax\ifmmode#1\else$#1$\fi}

\newcommand{\sw}{\mathswitch {s_{\scrs\PW}}}
\newcommand{\cw}{\mathswitch {c_{\scrs\PW}}}

\newcommand{\cew}{C^{\ew}}

\def\eg{e.g.\ }

\newcommand{\ord}{{\cal O}}
\newcommand{\Oa}{\mathswitch{{\cal{O}}(\alpha)}}

\newcommand{\ew}{\mathrm{ew}}

\newcommand{\rL}{\mathrm{L}}
\newcommand{\rT}{{\mathrm{T}}}
\newcommand{\rS}{{\mathrm{S}}}
\newcommand{\rd}{{\mathrm{d}}}

\newcommand{\pT}{p_{\mathrm{T}}}
\newcommand{\pTcut}{p_{\mathrm{T}}^{\mathrm{cut}}}

\newcommand{\M}{{\cal {M}}}
\newcommand{\calL}{{\cal L}}

\newcommand{\NLLa}{\stackrel{\mathrm{NLL}}{=}}

\newcommand{\shat}{{\hat s}}
\newcommand{\that}{{\hat t}}
\newcommand{\uhat}{{\hat u}}
\newcommand{\rhat}{{\hat r}}

\newcommand{\rar}{{\rightarrow}}

\newcommand{\MSBAR}{\overline{\mathrm{MS}}}

\def\draftdate{\relax}
\def\mpar#1{\relax}
\def\mua{\relax}
\def\mda{\relax}
\def\mla{\relax}
\def\draft{
\def\thtystars{******************************}
\def\sixtystars{\thtystars\thtystars}
\typeout{}
\typeout{\sixtystars**}
\typeout{* Draft mode!
         For final version remove \protect\draft\space in source file *}
\typeout{\sixtystars**}
\typeout{}
\def\draftdate{\today}
\def\mua{\marginpar[\boldmath\hfil$\uparrow$]%
                   {\boldmath$\uparrow$\hfil}%
                    \typeout{marginpar: $\uparrow$}\ignorespaces}
\def\mda{\marginpar[\boldmath\hfil$\downarrow$]%
                   {\boldmath$\downarrow$\hfil}%
                    \typeout{marginpar: $\downarrow$}\ignorespaces}
\def\mla{\marginpar[\boldmath\hfil$\rightarrow$]%
                   {\boldmath$\leftarrow $\hfil}%
                    \typeout{marginpar: $\leftrightarrow$}\ignorespaces}
\def\Mua{\marginpar[\boldmath\hfil$\Uparrow$]%
                   {\boldmath$\Uparrow$\hfil}%
                    \typeout{marginpar: $\Uparrow$}\ignorespaces}
\def\Mda{\marginpar[\boldmath\hfil$\Downarrow$]%
                   {\boldmath$\Downarrow$\hfil}%
                    \typeout{marginpar: $\Downarrow$}\ignorespaces}
\def\Mla{\marginpar[\boldmath\hfil$\Rightarrow$]%
                   {\boldmath$\Leftarrow $\hfil}%
                    \typeout{marginpar: $\Leftrightarrow$}\ignorespaces}
\def\mpar##1{\marginpar{\hbadness10000%
                      \sloppy\hfuzz10pt\boldmath\bf##1}%
                      \typeout{marginpar: ##1}\ignorespaces}

\overfullrule 5pt
\oddsidemargin -15mm
\marginparwidth 29mm
}

\begin{document}

\thispagestyle{empty}
\def\thefootnote{\fnsymbol{footnote}}
\setcounter{footnote}{1}
\null
\draftdate
\hfill   TTP07-07\\
\strut\hfill  SFB/CPP-07-12\\
\strut\hfill  MPP-2007-35\\
\strut\hfill  DESY 07-041\\
\strut\hfill hep-ph/0703283
\vskip 0cm
\vfill
\begin{center}
{\Large \bf
Electroweak corrections to  large transverse momentum
production of $W$ bosons at the LHC
\par}

%\bigskip
%\bigskip
%\bigskip
\bigskip

{\large \sc
Johann~H.~K\"uhn$^a$,
A.~Kulesza$^b$,
S.~Pozzorini$^c$,
M.~Schulze$^a$}

%\bigskip
\bigskip

\begin{it}

$^a$Institut f\"ur Theoretische Teilchenphysik, 
Universit\"at Karlsruhe,\\
D-76128 Karlsruhe, Germany\\
\bigskip
$^b$
Deutsches Elektronen-Synchrotron DESY, Notkestrasse 85,\\
D--22607 Hamburg, Germany \\
\bigskip
$^c$Max-Planck-Institut f\"ur Physik, F\"ohringer Ring 6,\\
D--80805 Munich, Germany

\end{it}

\bigskip
\end{center}

{\bf Abstract:} \par 
To match the precision of present and future measurements 
of $W$-boson production at hadron colliders,
electroweak radiative corrections must be included
in the theory predictions.
In this paper we consider their effect on the 
transverse momentum ($p_\rT$) distribution of $W$ bosons, 
with emphasis on large $p_\rT$.
We evaluate the full electroweak $\ord(\alpha)$ corrections 
to the process 
$pp\to W j$ including 
virtual and real photonic contributions.
We also provide compact approximate expressions which are valid 
in the high-energy region, where the electroweak corrections are strongly 
enhanced by logarithms of $\shat/M_W^2$.
These expressions include quadratic and single logarithms 
at one loop as well as quartic and triple logarithms at two loops.
Numerical results are presented for proton-proton
collisions at 14 TeV.
The corrections are negative and their size increases with $p_\rT$.
At the LHC, where transverse momenta of 2 TeV or more can be reached,
the one- and two-loop corrections amount up to 
$-40\%$ and $+10\%$, respectively.

\par
\vskip 1cm
\noindent
March 2007 
\par
%\null
\setcounter{page}{0}
\clearpage
\def\thefootnote{\arabic{footnote}}
\setcounter{footnote}{0}

\newpage

\section{Introduction}

The study of gauge-boson production has been among the primary 
goals of hadron colliders, starting with the discovery of the $W$ 
and $Z$ bosons more than two decades ago 
\cite{Arnison:1983rp}.
The investigation of the production dynamics, strictly predicted by 
the electroweak theory, constitutes one of the important tests
of the Standard Model. 
Differential distributions of gauge bosons, in rapidity as well
as in transverse momentum ($p_\rT$), 
have always been the subject of
theoretical and experimental studies.

At large $\pT$ the final state of
the leading-order 
(LO) process consists of 
an electroweak gauge boson
%a $W$ or $Z$ boson 
plus one recoiling jet.
The high center-of-mass energy at the Large Hadron Collider (LHC) in
combination
with the enormous luminosity will allow to explore
parton-parton
scattering up to 
energies of several TeV 
and correspondingly production of gauge
bosons with transverse momenta up to \mbox{2 TeV}
or even beyond. 
In this energy range the electroweak
corrections from virtual weak-boson exchange 
are strongly enhanced, with the dominant terms 
in $L$-loop approximation being 
leading logarithms (LL) of the form 
$\alpha^L\log^{2L}(\hat{s}/M_W^2)$,
next-to-leading logarithms (NLL) of the form 
$\alpha^L\log^{2L-1}(\hat{s}/M_W^2)$, and so on.
These corrections, also known as electroweak Sudakov logarithms, may
well amount to several tens of percent~\cite{Kuhn:2000nn,Fadin:2000bq,Dittmaier:2001ay,Maina:2004rb,Kuhn:2004em,Baur:2006sn}.
A recent survey of the literature on 
electroweak Sudakov logarithms
can be found in \citere{Denner:2006jr}.
Specifically, the electroweak corrections to the $\pT$ distribution of
photons and $Z$ bosons at hadron colliders were studied in \citeres{Maina:2004rb,Kuhn:2004em}.
In \citere{Kuhn:2004em}, it was found that at transverse momenta of $\ord(1\TeV)$
the dominant two-loop contributions to these reactions
amount to several percent and 
must be included
to
match the precision of the LHC experiments.

In this paper we study the electroweak corrections to the hadronic production of $W$ bosons at large $p_\rT$.
In contrast to the case of $Z$ and $\gamma$ production, the contributions
from virtual and real photons cannot be separated from the purely weak corrections and 
will thus be included in our analysis.

The partonic 
reactions $\bar{q}q'\rightarrow W^\pm g\, (\gamma)$, 
$q'g\rightarrow W^\pm q\, (\gamma)$ and
{$\bar{q}g\rightarrow W^\pm \bar{q}'\, (\gamma)$}  
with $q=u,d,s,c,b$ are considered. 
All of them are, however, trivially related 
by CP- and crossing-symmetry relations.
Quark-mass effects are neglected throughout, 
which allows to incorporate the effect of quark mixing 
through a simple redefinition of parton distribution functions
(see~\refse{se:kinematics}). 
The calculation of the 
virtual corrections is described in~\refse{se:virt}. In this Section
we also 
present compact analytic expressions for the high-energy behaviour 
of the  corrections which include quadratic and linear logarithms 
at one loop as well as quartic and triple logarithms at two loops.
The calculation of the real corrections is performed using the dipole subtraction 
formalism \cite{Dittmaier:1999mb,Catani:1996vz,Catani:2002hc}, as
described in~\refse{se:real}.
After convolution with parton distribution functions, we obtain
radiatively corrected predictions for $\pT$ distributions
of $W$ bosons at the LHC, presented in~\refse{se:numerics}.
Concerning perturbative QCD, our predictions are based on the lowest
order.
To obtain realistic absolute cross sections,
higher-order QCD corrections \cite{QCDcorr} must be included.
However, the relative rates for $W^+$, $W^-$ and $Z$
production are expected to be more stable against 
QCD effects. Therefore, in \refse{se:numerics}
we also study the impact of the electroweak corrections on these ratios.

\section{Lowest order and kinematics}
\label{se:kinematics}%\refse{se:kinematics}

The  $\pT$ distribution of $W$ bosons in the reactions 
$h_1 h_2 \to W^\pm j$ 
is given by 
\newcommand{\pdf}[4]{f_{#1,#2}(#3,#4)}
\newcommand{\pdfmod}[4]{\tilde f_{#1,#2}(#3,#4)}
\beq
\label{hadroniccs}%\refeq{hadroniccs}
\frac{\rd \si^{h_1 h_2}}{\rd \pT}=
\sum_{i,j,k}\int_0^1\rd x_1 \int_0^1\rd x_2
\;\theta(x_1 x_2-\hat\tau_{\rm min})
\pdf{h_1}{i}{x_1}{\mu^2}
\pdf{h_2}{j}{x_2}{\mu^2}
\frac{\rd \hat{\si}^{i j\to W^\pm k}}{\rd \pT}
,  
\eeq
where 
$\hat \tau_{\rm min} =(\pT+m_\rT)^2/s$,
$m_{\rT}=\sqrt{\pT^2+ M_W^2}$  
and $\sqrt{s}$ is the collider energy.
The indices  $i, j$  denote initial-state partons %($q, \bar q, g$) 
and
$f_{h_1,i}(x,\mu^2)$, $f_{h_2,j}(x,\mu^2)$
are the corresponding parton distribution functions.
$\hat {\si}^{ij\to W^\pm k}$ is the partonic cross section for 
the subprocess $i j \to W^\pm k$ and the sum 
runs over all $i,j,k$ combinations corresponding to the subprocesses
\beqar\label{processes1}%\refeq{processes1}
&&\bar u_m d_n \to W^- g,\quad
d_n\bar u_m\to W^- g,\quad
g d_n \to W^-  u_m,\quad
d_n g \to W^-  u_m,\quad
\nl&& \bar u_m g  \to W^-  \bar d_n,\quad
g \bar u_m   \to W^-  \bar d_n
,
\eeqar
for $W^-$ production, and similarly for $W^+$ production. 
The dependence of the partonic cross sections on the flavour indices $m,n$
amounts to an overall factor $|V_{u_md_n}|^2$. This factor can be easily absorbed 
in the parton distribution functions
of down-type quarks by redefining 
\newcommand{\parbar}[2]{{#1_#2}}
\beqar
{f}_{h,\parbar{d}{m}}%(x,\mu^2)
&\to&
\sum_{n=1}^3|V_{u_m d_n}|^2 f_{h,\parbar{d}{n}}%(x,\mu^2) 
,\qquad
{f}_{h,\parbar{\bar d}{m}}%(x,\mu^2)
\to
\sum_{n=1}^3|V_{u_m d_n}|^2 f_{h,\parbar{\bar d}{n}}%(x,\mu^2) 
.
\eeqar
The partonic cross sections can then be computed using
 a trivial CKM matrix $\de_{mn}$.
The Mandelstam variables for the subprocess $i j \to W^\pm k$  are defined
in the standard way
\beq
\shat=(p_i+ p_j)^2 
,\qquad 
\that=(p_i- p_W)^2 
,\qquad 
\uhat=(p_j-p_W)^2
.
\eeq
Momentum conservation implies $\shat+\that+\uhat=M_W^2$, and the invariants are related to the collider energy $s$ and the transverse momentum $p_\rT$ by $p_\rT^2=\that \uhat/\shat$ with $\shat=x_1 x_2 s$. 

The $\pT$ distribution 
for the unpolarized partonic subprocess $ij\to W^\pm k$
reads 
\beqar\label{partoniccs1}%\refeq{partoniccs1}
\frac{\rd \hat{\si}^{i j\to W^\pm k}}{\rd \pT}
&=&
\frac{\pT}{8\pi N_{i j}\shat|\that-\uhat|}
\left[
\overline{\sum}|\M^{i j\to W^\pm k}|^2+(\that\leftrightarrow \uhat)
\right]
,
\eeqar
where
$
\overline{\sum}=
\frac{1}{4}
\sum_{\mathrm{pol}} 
\sum_{\mathrm{col}} 
$
involves the sum over polarization and color as well as the 
factor $1/4$ for averaging over initial-state polarization. 
The factor $1/N_{i j}$ accounts for  the initial-state colour average.

The unpolarized squared matrix elements for all partonic processes 
relevant for $W^+$ and $W^-$ production 
are related by crossing- and CP-symmetry relations.
Thus the explicit computation of the unpolarized squared matrix element
 needs to be performed only once,
\eg for $\bar{u}d\to W^-g$. 
For this reaction, to lowest order in $\alpha$ and $\alpha_\rS$, we have
\beq\label{generalamplitude}%\refeq{generalamplitude}
\overline{\sum}|\M_{\rm Born}^{\bar{u}d\to W^- g}|^2=
32 \pi^2 \frac{\alpha}{\sw^2} \alpha_\rS 
\frac{\that^2+\uhat^2+2 M_W^2 \shat}{\that\uhat} 
,
\eeq
where $\sw=\sqrt{1-\cw^2}$ denotes the sine of the weak mixing angle.

\section{Virtual corrections}
\label{se:virt}%\refse{se:virt}

The one-loop diagrams were reduced to a minimal set
of coupling structures, standard matrix elements and scalar integrals
as in  \citere{Kuhn:2004em}.
The electroweak coupling constants were renormalized 
in the $G_\mu$-scheme, 
where  
$\alpha=\sqrt{2}\,G_\mu M_W^2\sw^2/\pi$ is expressed in terms of the
Fermi constant $G_\mu$
and
$\sw^2=1-M_W^2/M_Z^2$.
Soft and collinear singularities resulting from virtual photons
were regularized and combined with corresponding singularities from real photons as described in \refse{se:real}.
Complete analytic results for the one-loop corrections 
and their asymptotic behaviour will be provided in 
\citere{nextpaper}. 
The numerical evaluation and detailed cross checks 
were performed with two independent programs.
For the scalar loop integrals  we used the Fortran library 
 \cite{Beenakker:1988jr} and the {\tt FF} library~\cite{vanOldenborgh:1990yc}.

In the following, we present compact analytic expressions for the
one- and two-loop NLL contributions at high energy.
As in the case of $Z$ and $\gamma$ production  \cite{Kuhn:2004em},
the NLL terms are obtained from the Born result by multiplication with a global factor.
For the process  $\bar{u} d\to W^- g$ we have 
\newcommand{\logar}[2]{\mathrm{L}^{#1}_{#2}}
\beq\label{generalamplitudetwo}%\refeq{generalamplitudetwo}
\overline{\sum}|\M^{\bar{u}d\to W^- g}|^2
=
\overline{\sum}|\M_{\mathrm{Born}}^{\bar{u}d\to W^- g}|^2
\left[
1
+\left(\frac{\alpha}{2\pi}\right)A^{(1)}
+\left(\frac{\alpha}{2\pi}\right)^2 A^{(2)}
\right]
.
\eeq
At one loop, the NLL part consists of double- and single-logarithmic
terms and reads
\beq\label{oneloopresult}%\refeq{oneloopresult}
A^{(1)}\NLLa - 
\cew_{q_\rL}\left(\logar{2}{\shat}-3\logar{}{\shat}\right)
-
\frac{C_{\mathrm{A}}}{2 \sw^2}
\left(\logar{2}{\that}+\logar{2}{\uhat}-\logar{2}{\shat}\right)
.
\eeq
Here $\cew_{q_\rL}=Y_{q_\rL}^2/(4\cw^2)+C_{\mathrm{F}}/\sw^2$,
$C_{\mathrm{F}}=3/4$, $C_{\mathrm{A}}=2$, and
$\logar{}{\rhat}=\log(|\rhat|/ M_W^2)$.
At two loops we obtain
\beqar\label{twolooplogs}%\refeq{twolooplogs}
A^{(2)}&\NLLa&
\frac{1}{2}
\left(\cew_{q_\rL}+\frac{C_{\mathrm{A}}}{2\sw^2}\right)
\Biggl[\cew_{q_\rL}\left(\logar{4}{\shat}-6\logar{3}{\shat}\right)
+
\frac{C_{\mathrm{A}}}{2\sw^2}
\left(\logar{4}{\that}+\logar{4}{\uhat}-\logar{4}{\shat}\right) 
\Biggr]
\nl&&{}
+\frac{1}{6}
\Biggl[
\frac{b_1}{\cw^2}\left(\frac{Y_{q_\rL}}{2}\right)^2
+\frac{b_2}{\sw^2} \left(
C_{\mathrm{F}}
+\frac{C_{\mathrm{A}}}{2}\right)
\Biggr]\logar{3}{\shat}
,
\eeqar
where $b_1=-41/(6\cw^2)$ and $b_2=19/(6\sw^2)$.
The NLL results \refeq{oneloopresult}--\refeq{twolooplogs}
include the full electroweak corrections in the $M_Z=M_W$ approximation.
The photonic singularities are separated using 
the fictitious photon 
mass%
\footnote{
For a discussion of this prescription, as well as for details concerning the
implementation of the angular-dependent part of the NLL terms,
we refer to \citere{Kuhn:2004em}.} 
$\la=M_W$.
Eq.~\refeq{twolooplogs} has been derived from the general 
results 
for leading- and next-to-leading electroweak two-loop logarithms
in \citere{Fadin:2000bq}.

\section{Real radiation}
\label{se:real}%\refse{se:real}

Soft and collinear singularities from 
real radiation are treated using the dipole subtraction formalism~\cite{Dittmaier:1999mb,Catani:1996vz,Catani:2002hc}.
Within this framework infrared singularities 
of a squared amplitude for real radiation are subtracted 
by an auxiliary function that
has the same singular behaviour.
This ensures that the phase space integral of the difference
is a finite quantity and the integration can be performed numerically.
The analytical result for the integral of the auxiliary function over
the subspace of a radiated particle is then added to 
the result for virtual corrections. The
singularities of the virtual corrections cancel against those of the 
integrated subtraction part.   

The algorithms for constructing the auxiliary subtraction function and
its integrated counterpart have been developed both for the case
of photon radiation off massless or massive
fermions~\cite{Dittmaier:1999mb} and QCD radiation off massless~\cite{Catani:1996vz} or massive
partons~\cite{Catani:2002hc}.  The
latter~\cite{Catani:1996vz,Catani:2002hc} approach can be easily
adopted to the case of photon radiation. It employs dimensional
regularization to regularize soft
and collinear singularities while the former
approach~\cite{Dittmaier:1999mb} introduces small photon and fermion
masses. We have performed independent calculations of the 
virtual and real corrections within the two regularization schemes and verified that
the soft and collinear (apart from 
initial-state and final-state
collinear) singularities cancel. We also checked that the numerical results 
obtained within the massive and massless 
regularization scheme are in agreement.
In both approaches we use expressions 
for the emission off a massive fermion to describe the emission off a $W$ boson, 
since 
only soft singularities are present in this case and they depend only on 
the charge of the external particle and not on its spin. 

We restrict our analysis to $W$-boson production at high $\pT$  by  
cutting away events with low $\pT(W)$. Additionally, in order 
to avoid soft-gluon singularities which can arise as
a side-effect of hard-photon radiation, we introduce a cut on 
$\pT(\mathrm{jet})$.

The remaining initial-state collinear singularities  
are absorbed into the 
definition of parton distribution functions (PDFs). We choose to
perform calculations in the
$\MSBAR$ factorization scheme with the scale $M_W^2$. 
The photon-induced 
processes are not included in our calculations
since they are expected to be suppressed due to the size of the 
electromagnetic coupling%
\footnote{
In order to consistently include $\Oa$ corrections in a
calculation of a hadronic cross section, PDFs that are used in the
calculation need to take into account QED effects. Such PDF analysis
has been performed in~\cite{Martin:2004dh} and the $\Oa$ effects are
known to be small, both concerning the change in the quark
distribution functions (below ${\cal O}(1\%)$~\cite{Roth:2004ti}) and
the size of the photon distribution function. These effects are below
typical uncertainty of hadronic processes.  Moreover, the currently
available PDFs incorporating $\Oa$ corrections,
MRST2004QED~\cite{Martin:2004dh}, include QCD effects at the NLO in
$\al_\rS$.  Since our calculations are of the lowest order in QCD, and
QED effects on PDFs are estimated to be small, we choose to use a LO
QCD PDF set without QED corrections {incorporated, rather than
MRST2004QED.
\vspace{-5mm}}
}.

For the gluon-induced subprocesses, final-state collinear
singularities are also present. These singularities from collinear 
photon-quark configurations
are avoided by recombining photons and quarks in the collinear region.
In practice, we define the separation variable
\beq
R(q,\gamma)=\sqrt{(\eta_q-\eta_\gamma)^2+(\phi_q-\phi_\gamma)^2},
\eeq  
where $\eta_i$ is the pseudo-rapidity
and $\phi_i$ is the azimuthal angle of a particle $i$. 
If $R(q,\gamma)<R_{\mathrm{sep}}$, then the photon and quark
momenta are recombined into an 
effective momentum $\pT(\mathrm{jet})=\pT(q+\gamma)$ which is
subjected to the aforementioned cut.

\section{Predictions for the hadronic $W$ production at high transverse
  momentum}
\label{se:numerics}%\refse{se:numerics}
For the numerical evaluation of the corrections we use the following 
input parameters:
 $G_\mu=1.16637\times10^{-5}\GeV^{-2}$,
$M_W=80.39\GeV$,
$M_Z=91.19\GeV$,
$M_t=171.4\GeV$,
$M_H=120\GeV$.
Light-fermion and $b$-quark masses are neglected.
The hadronic cross sections are obtained using LO MRST2001 PDFs~\cite{Martin:2002dr}.
We choose $\mu^2=\pT^2(W)$ as the factorization scale
and, similarly, as the scale at which the running strong coupling constant
is evaluated. We also adopt, in agreement with the value used in
the PDF analysis, the value 
$\al_\rS(M_Z^2)=0.13$ and use the one-loop running expression for 
$\al_\rS(\mu^2)$. For the numerical values of elements in the CKM quark
mixing matrix we refer to~\cite{Eidelman:2004wy}. 
The statistical accuracy of the 
$W j$ cross section at LHC 
is estimated using the integrated luminosity
$\mathcal{L}=300 \mathrm{fb}^{-1}$,
and the branching ratio BR($W\to e\nu_e+\mu\nu_\mu$)=$2/9$.

We apply the 
%following values of the $\pT$ 
cuts $\pT(W)>100 \GeV$ and
$\pT(\mathrm{jet})>100 \GeV$. The value of the separation parameter
below which the recombination procedure is applied is taken to be 
$R_{\mathrm{sep}}=0.1$.

\begin{figure}[]
\vspace{2mm}
\begin{center}
\epsfig{file=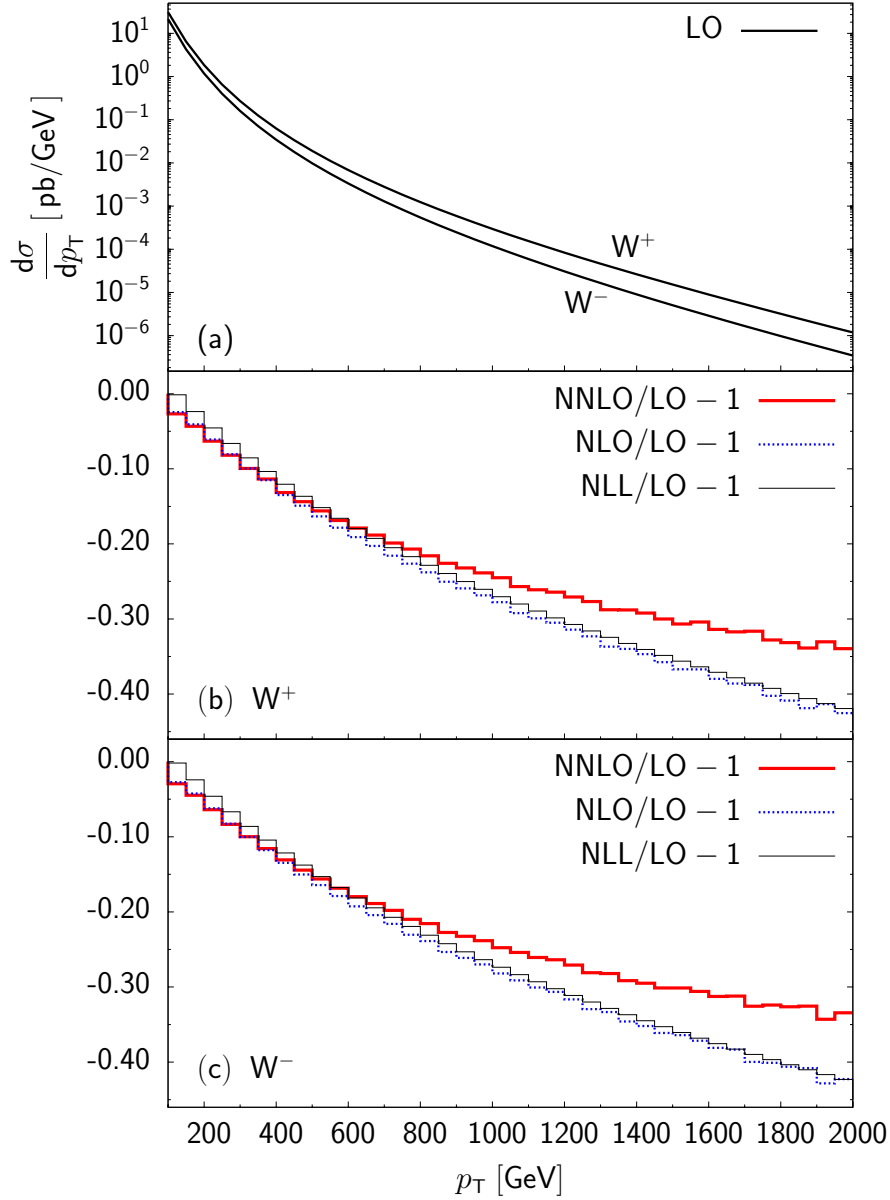, angle=0, width=11.5cm}
\end{center}
\vspace{-2mm}
\caption{Transverse momentum distributions for $W$-boson production at
the LHC:
LO predictions (a) 
and relative electroweak corrections
for $W^+$ (b) and $W^-$ (c) production
in NLO (dotted), one-loop NLL (thin solid) and NNLO (thick solid) approximation.
}
\label{fig:ptwplusminuslhc}
\end{figure}

The transverse momentum distributions 
in  LO approximation 
for $p p \rightarrow W^+ j$  and $pp\rightarrow  W^- j$
at $\sqrt{s}=14 \TeV$
are shown in \reffi{fig:ptwplusminuslhc}a. 
In~\reffi{fig:ptwplusminuslhc}b 
we plot the 
full $\ord(\alpha)$ electroweak (NLO), 
one-loop next-to-leading logarithmic (NLL) and next-to-next-to-leading order (NNLO)
corrections for $W^+$ production.
The NNLO corrections are defined as the NLO plus the two-loop NLL 
contributions \refeq{twolooplogs}.
As expected, 
the NLO corrections 
increase significantly with $\pT$.
They result in a negative contribution ranging from
$-15\%$ at $\pT=500 \GeV$ to $-42\%$ at $\pT=2 \TeV$. 
The one-loop NLL approximation \refeq{oneloopresult}
is in good agreement 
(at the 1-2\%  level) with the full NLO result.
The difference between NLO and NNLO curves is significant.
As can be seen from the plot the two-loop terms are positive and amount to  
%$+1\%$ at  $\pT=500\GeV$ 
$+3\%$ at  $\pT=1\TeV$  and 
$+9\%$ at  $\pT=2\TeV$. 
The behaviour of the relative corrections for $W^-$ production (\reffi{fig:ptwplusminuslhc}c) is qualitatively and quantitatively very similar.

%%%%%%%%%%%%%%%%%%%%%%%%%%%%%%

\begin{figure}[]
\vspace*{2mm}
  \begin{center}
\epsfig{file=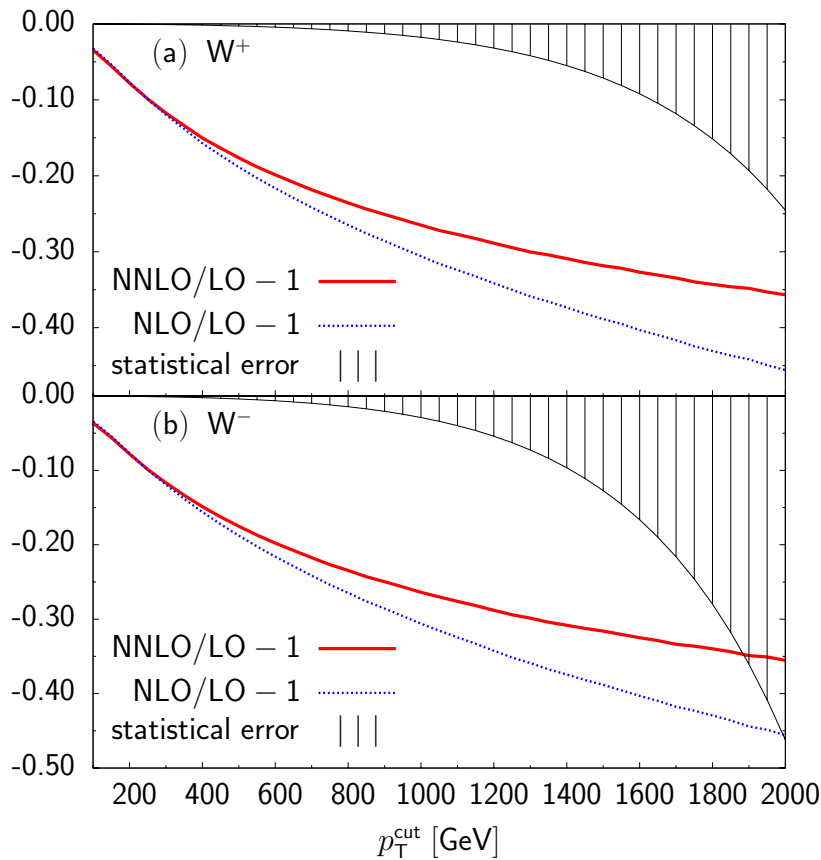, angle=0, width=10.5cm}
\end{center}
\vspace*{-2mm}
\caption{Unpolarized integrated cross section as a function of
$\pTcut(W)$ for $W^+$ (a) and $W^-$ (b) production: estimated
statistical error (shaded area) and relative electroweak corrections
in NLO (dotted) and NNLO (solid) approximation.  }
\label{fig:cswplusminuslhc}
\end{figure}

To underline the relevance of the large electroweak corrections
at the LHC, 
in \reffi{fig:cswplusminuslhc}a and  \reffi{fig:cswplusminuslhc}b we present 
the relative NLO and NNLO 
corrections to the $W^+$ and $W^-$ cross sections
integrated over $\pT$ starting from $\pT = \pTcut$, as a function of
$\pTcut$. This is compared with the statistical error, 
estimated as
$\Delta \si_{\rm stat} / \si = 1 /\sqrt N$ with $N= \calL \times {\rm
  BR} \times \si_{\rm LO}$ where the BR accounts for
the $e \bar{\nu_e}$ and $\mu \bar{\nu_\mu}$ decay modes
(for this estimate we ignore experimental efficiencies and cuts).
It is clear that the size of 
the NLO corrections is much bigger then the statistical error.
Also the difference between NNLO and NLO corrections, due to
two-loop logarithmic effects, is significant.
In terms of  the estimated statistical error, these
two-loop contributions amount to 
1--3 standard deviations 
for $p_\rT=\ord(1\TeV)$.
%%%%%%%%%%%%%%%%%%%%%%%%%%%%%%
%%%%%%%%%%%%%%%%%%%%%%%%%%%%%%%

\begin{figure}[]
\vspace*{2mm}
  \begin{center}
\epsfig{file=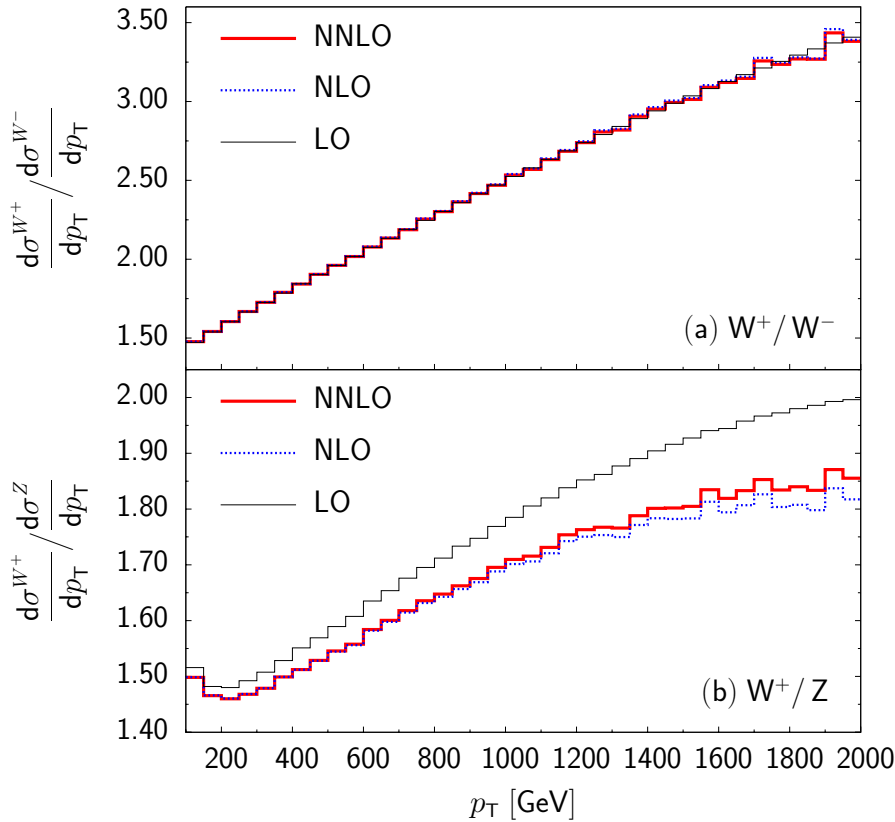, angle=0, width=11.5cm}
\end{center}
\vspace*{-2mm}
\caption{Ratio of the transverse momentum distributions
for the processes (a) $p p\rar W^+ j$ and  $p p\rar W^- j$ and 
(b) $p p\rar W^+ j$ and  $p  p\rar Z j$ 
at $\sqrt{s}=14 \TeV$:
LO (thin solid), NLO (dotted) and NNLO (thick solid) predictions.
}
\label{fig:ratiowplusminuszlhc}
\end{figure}

Finally let us discuss the ratios of the $p_\rT$ distributions for
 $W^+$ , $W^-$ and $Z$ bosons \cite{Kuhn:2004em}.
In contrast to the distributions themselves, these ratios are expected to be 
relatively insensitive to QCD corrections and theoretical uncertainties associated with $\alpha_\rS$ and PDFs. Therefore they are good candidates for precision measurements.
The $W^+/W^-$ ratio is presented in \reffi{fig:ratiowplusminuszlhc}a.
In the considered $p_\rT$ range  the LO value increases from
1.5 to 3.5. As already observed, 
the (relative) electroweak corrections to the 
$W^+$ and $W^-$ production processes 
are almost identical.
As a consequence the LO, NLO and NNLO curves in  \reffi{fig:ratiowplusminuszlhc}a overlap.
In contrast, 
the impact of the electroweak corrections 
on the $W^+/Z$ ratio (\reffi{fig:ratiowplusminuszlhc}b) 
is clearly visible.
Here the LO 
prediction, ranging from 
1.5 to 2,
receives corrections that grow with $p_\rT$ and 
amount to 5-10\% for $p_\rT \ge 1\TeV$.
%%%%%%%%%%%%%%%%%%%%%%%%%%%%%%
%%%%%%%%%%%%%%%%%%%%%%%%%%%%%%
%

\section{Summary}
In this work we evaluated the electroweak corrections to 
large transverse momentum production of
$W$ bosons at the LHC. The contributions
from real and virtual photons cannot be separated in a gauge
invariant manner from purely weak corrections
and were thus included in our analysis. Soft and collinear
singularities were regulated by introducing a small quark mass and a
small photon mass and, alternatively, by using dimensional
regularization.
The real photon radiation was evaluated using the
dipole formalism. The  $\ord(\alpha)$  corrections lead to a reduction 
of the cross section by about 
15\% at transverse momenta of 500~GeV  
and reach more than $-40\%$ at 2~TeV. 
We also derived a compact (NLL) approximation which includes the
quadratic and linear logarithms and which  gives a good 
description of the full   $\ord(\alpha)$
result with an accuracy of about 
1--2\%.
Considering the large event rate at the LHC,
leading to a fairly good statistical precision
even at transverse momenta up to 2 TeV,
we evaluated also the dominant (NLL) two-loop terms. 
In the high-$p_\rT$ region, these two-loop effects increase the cross
section by 5-10\% and thus become of importance in precision studies.
We also studied the relative rates for $W^+$, $W^-$ and
$Z$ production, which are expected to be stable with respect to
QCD effects. 
The electroweak corrections cancel almost completely in the $W^+/W^-$
ratio.  Instead, their impact on the $W^+/Z$ ratio is significant and
amounts to several percent for $p_\rT \ge 1\TeV$.

\subsection*{Acknowledgements}
We would like to thank S.~Dittmaier, B.~J\"ager and P.~Uwer for
helpful discussions.
This work was supported in part by
BMBF Grant No.\ 05HT4VKA/3, the Sonderforschungsbereich Transregio 9
and the DFG
Graduiertenkolleg ``Hochenergiephysik und Teilchenastrophysik''.

\end{document}